\documentclass[3p,times,procedia]{elsarticle}
\usepackage{nupha_ecrc}

\newcommand{\dd}{\ensuremath{\mathrm{d}}}

\volume{00}

\firstpage{1}

\journalname{Nuclear Physics A}

\runauth{}

\jid{nupha}

\jnltitlelogo{Nuclear Physics A}

\usepackage{amssymb}

\usepackage[figuresright]{rotating}

\begin{document}

\begin{frontmatter}

\dochead{XXVIIth International Conference on Ultrarelativistic Nucleus-Nucleus Collisions\\ (Quark Matter 2018)}

\title{Quarkonium: a theory overview}

\author{Elena G. Ferreiro}

\address{Departamento de F{\'\i}sica de Part{\'\i}culas and IGFAE, Universidade de Santiago de Compostela, 15782 Santiago de Compostela, Spain\\
Laboratoire Leprince-Ringuet, Ecole polytechnique, CNRS/IN2P3, Universit\'e Paris-Saclay, Palaiseau, France}

\begin{abstract}
Quarkonium has long been proposed as one of the golden probes to identify the
phase transition from confined hadronic matter to
the deconfined quark-gluon plasma in heavy-ion collisions. Since then, we have achieved a better understanding, not only about the propagation of quarkonium in-medium, but also about its production.
Recent theoretical developments in our comprehension 
of the quarkonium production mechanism in proton-proton collisions and on the propagation of quarkonia in proton-nucleus
and nucleus-nucleus collisions are
reviewed and discussed.
\end{abstract}


\end{frontmatter}

\section{Introduction}
\label{Intro}
Since the discovery of the $J/\psi$ charmonium meson in 1974 followed by the $\Upsilon$ bottomonium meson in 1977, the atention dedicated to quarkonia 
has not ceased to rise, becoming even more prominent in the last years. On one side, quarkonium bound states offer a solid ground to probe Quantum Chromodynamics (QCD), 
due to the high scale provided by the large mass of the heavy quarks. On the other side, the properties of production and absorption of quarkonium in a nuclear medium provide quantitative inputs for the study of the
matter created in ultrarelativistic heavy-ion collisions. In particular, quarkonium can be used as a probe of the properties of the medium created in these collisions, such as the strength of the interactions and the possible thermalisation.

From my point of view, our inquiry on quarkonium should be organised starting
from the production mechanisms in proton-proton collisions, followed by the modification induced by the proton-nucleus collisions, to finish by the investigation of the quarkonium properties in the hot and dense strongly-interacting quark-gluon Plasma (QGP) in nucleus-nucleus collisions.

Quarkonium production in proton-proton collisions provides an important tool to improve our understanding of various aspects of QCD. While the production of the heavy-quark pair is to be taken as perturbative, its hadronisation 
to form a quarkonium bound state is non-perturbative as it involves long distances and soft momentum scales. Thus, a careful study of heavy-flavour production within different theoretical models and their comparison to experimental data provides relevant testing grounds for both perturbative and non-perturbative aspects of QCD calculations and for the validity of QCD factorisation for the pair production.

On the other side, in nucleus-nucleus collisions, open and hidden heavy-flavour productions constitute sensitive probes of the hot strongly-interacting medium, since hard scattering processes take place in the early stage of the collision 
and since they subsequently interact with the hot and dense medium out of which they escape.
Moreover, disentangling the hot medium-induced effects and relating them to its properties requires an accurate study of the so-called cold nuclear matter (CNM) effects.
Those effects, present in proton-nucleus collisions, can
modify the production of heavy quarks in nuclear collisions with respect to proton-proton collisions, without invoking the creation of a QGP. They are obviously relevant if one wishes to use quarkonia as probes of a QGP formation, since they offer the natural baseline that needs to be taken into account.

\section{Quarkonium production in proton-proton collisions}
\label{pp}
The theoretical study of quarkonium production processes involves both perturbative and non-perturbative aspects
of QCD. On one side, the production of the heavy-quark pair, $Q\bar{Q}$ , which will subsequently form the quarkonium, is
expected to be perturbative since it involves momentum transfers at least as large as the mass of the considered heavy
quark. On the other side, the evolution of the
$Q\bar{Q}$ pair into the physical quarkonium state is non-perturbative, over long distances, with typical momentum scales
such as the momentum of the heavy-quarks in the bound-state rest frame $m_Qv $  and their binding energy $m_Q v^2$, $v$  being
the typical velocity of the heavy quark or antiquark in the quarkonium rest frame ($v^2=0.3$ for the charmonium and
0.1 for the bottomonium).
In nearly all the models or production mechanisms discussed nowadays, the idea of a factorisation between the pair
production and its binding is introduced. Various approaches differ essentially in the treatment of the hadronisation,
and some of them also introduce new ingredients in the description of the heavy-quark-pair production. In the
following, I briefly describe three of them which can be distinguished in their treatment of the non-perturbative
part: 
 the Colour-Singlet Model (CSM), the Colour-Octet Mechanism (COM) --both encompassed in the Non-Relativistic QCD (NRQCD) effective theory--
 and the Colour-Evaporation Model (CEM). 
 I will also present the last developments on the precedent approaches.

Historically, the first model to describe quarkonium production, the CSM \cite{Einhorn:1975ua,Chang:1979nn,Baier:1981uk}, relies on the assumption
that the quantum state of the pair does not evolve between its production and its hadronisation, neither in spin, nor in colour. 
The partonic cross section for quarkonium production involves that for the production of a heavy-quark pair with zero relative velocity $v$ 
in a colour-singlet state and in the same angular momentum and spin state as that of the to-be produced quarkonium. 

In in the mid 90's, the CDF analyses of direct $J/\psi$ and $\psi$(2S) production at $\sqrt{s}=1.8$ TeV \cite{Abe:1997jz}
shown that the measured rates were more than an order of magnitude larger than leading order (LO) CSM calculations.
These discrepancies motivated  new theoretical investigations on quarkonium hadroproduction, leading to the NRQCD factorisation framework \cite{Bodwin:1992ye}.
In this approach, quarkonium production can also proceed via creation of colour-octet $Q\bar{Q}$ pairs, present in higher-Fock states, whose effects are believed to be suppressed by powers of the relative $Q\bar{Q}$ velocity, $v$.
On the other hand, taking into account higher-order colour singlet contributions, i.e. next-to-leading order (NLO) corrections and approximate next-to-next-to-leading order (NNLO) contributions (known as NNLO$^*$)
can reduce the most severe discrepancies of the CSM, in particular for $p_T$ up to a couple of $m_Q$ \cite{Artoisenet:2008fc}. A full NNLO computation is however needed to confirm this trend.

Based on the principle of quark-hadron duality, in the CEM \cite{Fritzsch:1977ay,Halzen:1977rs} the production cross section of quarkonia is connected to the one of a $Q\bar{Q}$ pair in the region where its hadronisation is possible, i.e. between the kinematical threshold to produce a quark pair, $2m_Q$, and that to create the lightest open heavy-flavour hadron pair, $2m_H$. One parameter per quarkonium state, $F_Q$, related to a process-independent probability that the pair hadronises into this state, is introduced:
$
\sigma^{\rm (N)LO}_{\cal Q}= F_{\cal Q}\int_{2m_Q}^{2m_H}  
\frac{\dd\sigma_{Q\overline Q}^{\rm (N)LO}}{\dd m_{Q\overline Q}}\dd m_{Q\overline Q}$. 
While this model benefits from a successful phenomenology, the main issues are related to 
discrepancies in some transverse momentum spectra and 
the absence of predictions for polarisation observables. 

An improved version of the model, the ICEM, recently developed \cite{Ma:2016exq}, relates the average final state momentum, $p_{\psi}$, to the $c\bar{c}$ pair momentum, improving the results at high $p_T$. 
Moreover, by including an explicit charmonium mass dependence,  the predicted ratio of differential cross sections of two charmonium states is no longer $p_T$-independent. A
LO calculation of quarkonium polarisation in this framework can also be developed, showing longitudinal polarisation \cite{Cheung:2017osx}.

Another interesting development concerns the Color Glass Condensate (CGC) saturation model of gluon distributions in the proton when combined with NLO NRQCD matrix elements \cite{Ma:2014mri}.
This formalism has been applied to compute $J/\psi$ production at low $p_T$ in proton-proton collisions at collider energies. A very good description of the total cross sections, the rapidity distributions and the low momentum $p_T$  distributions  is obtained. Moreover, 
the results in this framework can be matched smoothly to NLO perturbative QCD results at high $p_T$, providing a unified description for quarkonium production in all phase space.

Furthermore, the study of new observables, as the production of quarkonium in high multiplicity $pp$ events, may help to quantified the impact of the initial parton saturation \cite{Ferreiro:2012fb,Ma:2018bax} that can be present in 
proton-proton collisions.

\section{Quarkonium production in proton-nucleus collisions}
\label{pA}
The suppression of the $J/\psi$ in ultrarelativistic nucleus-nucleus collisions has been proposed a long time ago as a signature of the formation of a QGP \cite{Matsui:1986dk}.
Results from proton-nucleus collisions, taken in the same kinematical conditions and properly extrapolated to nucleus-nucleus collisions, can be helpful to calibrate the contribution of the various nuclear effects --already present in $pA$ collisions and commonly known as cold nuclear matter (CNM) effects-- to the overall observed suppression.
Thus, proton-nucleus reactions are of crucial importance to correctly interpret nucleus-nucleus observations.

\vskip 0.3cm
Among the most important effects present in 
proton-nucleus reactions I would mention:
\begin{itemize}[--]
\vskip -0.3cm
\item the nuclear modification of the parton distribution functions in nuclei, usually known as shadowing;
\vskip -0.3cm
\item the phenomenon of gluon saturation in the low-$x$ regime, as implemented in the CGC;
\vskip -0.3cm
\item the process of energy loss in the propagation of quarks and gluons in the medium;
\vskip -0.3cm
\item the quarkonium-hadron interactions, as the nuclear absorption or the comover interaction.
\end{itemize}

Besides, $pA$ collisions can help to learn about different QCD features. They provide means to study:
\begin{itemize}[--]
\item the time-evolution of a $Q\bar{Q}$ pair and the dynamics of its hadronisation;
\item the quarkonium-production mechanisms, in particular the colour octet vs. singlet contributions;
\item the validity of collinear factorisation in the nuclear medium;
\end{itemize}
which can give us essential information about QCD at the interface between its perturbative and non-perturbative domain.
In the following, I briefly revisit the mentioned CNM effects and their state-of-the-art.

\subsection{Nuclear parton distribution functions}

In a collinear-factorisation framework, the nuclear effects on the parton dynamics can be described in terms of nuclear-modified parton distribution functions (nPDF).
As it is known since the 80's,
the nuclei are not just a simple collection of free nucleons, and the nuclear PDFs are not just equal to the sum of the PDFs of the ingredient nucleons.
Three regimes are usually identified for the nPDF to PDF ratio $R_i(x,Q^2)$, depending on the value of $x$:
\begin{enumerate}
\item a suppression ($R_i<1$), commonly referred to as shadowing, at small $x <10^{-2}$;
\item a possible enhancement ($R_i>1$), known as antishadowing, at intermediate values $10^{-2}< x<10^{-1}$, 
\item the EMC effect, a depletion that takes place at large $x>10^{-1}$.
\end{enumerate}

The modification of gluon densities in nuclei affects the yields of quarkonium production.
The $R_g(x,Q^2)$ parametrisations are determined by performing global fit analyses of lepton-nucleus and proton-nucleus data.
To compensate the lack of constraints, most recent global NLO analyses of nPDFs, nCTEQ15 \cite{Kovarik:2015cma} and EPPS16 \cite{Eskola:2016oht}, used RHIC pion and LHC jet data.
The gluon nPDFs in the small-$x$ regime are obtained by extrapolating nPDFs from larger $x$ region. Thus, different approaches with different parametrisations of the $x$-dependence of nPDFs at the initial scale can give different results.

The present LHC data on quarkonium are compatible with strong shadowing in the forward rapidity region, while hints of antishadowing appear in the backward region \cite{Lansberg:2016deg}. 
In fact, the precision of the current data is already much better than the nPDF uncertainties. 
It may offer an indication for constraining the gluon density in Pb \cite{Kusina:2017gkz}.

\subsection{Saturation in the Colour Glass Condensate approach}
One of the first approaches implementing the ideas of the CGC \cite{Fujii:2013gxa} for the quarkonium production --using CEM for the hadronisation processes-- showed predictions 
on $J/\psi$ suppression in $pA$ collisions clearly below data.
Since then, 
several improvements 
have been performed.
On one hand, 
using a similar approach but with an ameliorated treatment of the nuclear geometry and a different parametrisation of the dipole cross section,  Ducloe {\it et al.} \cite{Ducloue:2015gfa}
showed that the $J/\psi$ suppression in 
$p$Pb collisions was less pronounced.

Moreover, by implementing small-$x$ evolution in the NRQCD formalism \cite{Ma:2015sia}, as previously done for $pp$ collisions \cite{Ma:2014mri}, the results can be improved. 
The colour-channel dependence of the hadronisation process, which is simply ignored in the CEM, becomes important.
Depending on which one of the NRQCD channels dominates the $J/\psi$ production cross section in $p$Pb collisions at the LHC, the $J/\psi$ suppression predicted in this formalism agree with the current ALICE and LHCb measurements.

Current results from CGC formalism are thus on the same order as nPDF predictions. They are also very much widespread, as those from nPDFs approaches. It is necessary to wonder about the different physics at play. Note also that the saturation can only be apply in the forward rapidity region.

\subsection{Other nuclear effects: the nuclear absorption and the coherent energy loss}
The models discussed above accounts for initial-state effects. But it is legitimated to wonder about the presence of final-state effects. 
At low --SPS-- energies, the quarkonium bound states may be destroyed by inelastic scatterings with the nucleus remnants --the spectator nucleons-- 
if they are formed in the nuclear medium. This is the so-called nuclear absorption or nuclear break-up.
At higher --LHC-- energies, the quarkonium formation time is expected to be larger than the nucleus radius. In fact, this formation time has to be considered in the rest frame of the target
nucleus, which results in a boost of more than 
1000 at LHC mid rapidities. This implies that the quarkonium formation time will be larger than the Pb radius practically in the whole rapidity region, resulting in a negligible nuclear absorption at the LHC energies.

This has motivated the study of other effects. In particular, a coherent energy loss, scaling like the projectile energy, has been proposed \cite{Arleo:2010rb}. It can reproduce the data from fixed-target to LHC energies. In presence of such an energy loss, which arises from interferences between initial and final-state radiations, no additional break-up is needed. It is however not clear if gluon shadowing is needed when such an energy loss is at play.
Disentangling shadowing from coherent energy loss using the Drell-Yan process has recently be proposed \cite{Arleo:2015qiv}.

\subsection{The excited state puzzle: comovers at play}
Measurements of $J/\psi$ and $\psi(2S)$ production in proton(deuteron)-nucleus collisions, both at RHIC and LHC energies, 
show a stronger suppression of the excited state. The situation is the same in which concerns bottomonium: the excited $\Upsilon(2S)$ and $\Upsilon(3S)$ states suffer more suppression than the lower $\Upsilon(1S)$ state.

At low collision energies, such a relative suppression is naturally explained from final-state interactions with the remnants of the colliding lead nucleus, the already mentioned nuclear absorption.
Yet, at LHC energies, as explained above, the produced $Q\bar{Q}$ pair does not have the time to evolve into any physical state when it escapes the nuclear matter contained in the colliding nucleus. 
Consequently, one cannot invoke this mechanism to explain the relative suppression, neither can one invoke initial-state effects such as the modification of the nPDFs or the coherent energy loss, which are known to have a similar impact on the 
different states.
 
In fact, as for now, the approach which can successfully describe these results is the comover interaction model (CIM) \cite{Ferreiro:2012rq,Ferreiro:2014bia}. Such an approach accounts for final-state scatterings 
with comoving particles, i.e. particles with similar rapidities that happen to travel along with the $Q\bar{Q}$ pair. 
These interactions can occur over times long enough for the quarkonium to be formed; differences between the 1S, 2S and 3S suppression can thus happen.
The density of comovers is directly connected to
the particle multiplicity measured at that rapidity for the corresponding colliding system.
This implies that the suppression due to the comovers increases with the centrality of the collision and that, for asymmetric collisions as proton(deuteron)-nucleus, it will be stronger in the nucleus-going direction.

The adjustable parameter of the CIM is the cross section of quarkonium dissociation due to interactions with the comoving medium.
In the improved version of the CIM \cite{Ferreiro:2018wbd},  a generic formula for all the quarkonium states is proposed, which relates the energy distribution of the 
comovers with an effective temperature.

Other models, as the transport model \cite{Du:2018wsj} or the combined CGC+ICEM model \cite{Ma:2017rsu} incorporate final interactions similar in spirit to comover suppression.

\section{Quarkonium production in nucleus-nucleus collisions}
\label{pA}
As it is always recall, more than 30 years ago
Matsui and Satz \cite{Matsui:1986dk} propose the suppression of quarkonium as a signature of the formation of a QGP. The original idea is that,
due to the Debye-like screening of the $Q\bar{Q}$ potential in the deconfined medium, different quarkonium states will melt sequentially at different temperatures,
being the $J/\psi$ state melted above the deconfinement temperature.

This nice idea relies however on a time-independent notion of the melting process, based on purely real-model potentials, 
where popular candidates were the colour-singlet free energy,  the colour-singlet internal energy or even linear combinations of both quantities. 

An essential step took place in 2007, when Laine {\it et al.} \cite{Laine:2006ns} demonstrated
that the heavy-quark potential at first non-trivial order in resummed perturbation theory not only shows Debye screening but also features an imaginary part.
Thus, a static notion of a well-defined bound state above the deconfinement temperature becomes empty of meaning.

An intuitive way to interpret this fact is to consider that the real part of the finite-temperature potential between two heavy quarks captures the screened $Q\bar{Q}$ interaction, while the imaginary one captures dissociation
--originating from the Landau damping
following from the inelastic scatterings of the pair with the constituents of the deconfined medium or to colour rotations of the $Q\bar{Q}$ pair leading to its dissociation. 

Current efforts oriented to understand the dynamics of the in-medium evolution of the heavy-quark bound state
concerns lattice QCD calculations of complex in-medium heavy-quark  potential, as presented by A. Rothkopf  
[TUM-QCD collaboration] in this conference.
				
\subsection{Open quantum systems}
 To formalize this idea of decoherence in the language of quantum mechanics and to see how the imaginary part arises from the thermal fluctuations in the medium surrounding the $Q\bar{Q}$ pair, a description based on the theory of
open quantum systems (OQS) can be applied.

The OQS framework is useful to describe the evolution of an open system --the heavy quarks-- in contact with an environment --the rest of the particles-- from which we only know average properties.
To a good approximation, one can assume that the environment is not modified by the interaction with the system.

The real and imaginary parts of the in-medium 
heavy-quark potential can be related to the stochastic 
evolution of the in-medium wave function which
is perturbed by the thermal medium: the
stochastic term corresponds to the thermal noise, and the imaginary part of the potential is
related to the strengtht of the thermal noise. In this way, the noise provides an dynamical dissociation mechanism \cite{Kajimoto:2017rel,Akamatsu:2018xim}.

Other recent developments, some of them presented in this conference, also show that  the implementation of OQS theory leads to a
master equation with time evolution of the heavy-quark states. 
This is the case, for example, when the time evolution of the heavy-quark states in an expanding hot QCD medium is taking into account by implementing effective field theory --perturbative NRQCD-- in the framework 
of open quantum systems \cite{Brambilla:2017zei}. A Lindblad equation is obtained. This approach avoids classical approximations and the non-Abelian nature of QCD is taken into account through colour transitions.

In the same line, equations for the time evolution of the heavy-quark reduced-density matrix in a non-Abelian QGP are presented in \cite{Blaizot:2017ypk,Blaizot:2018oev}.
The relative motion of the heavy quarks is treated semi-classically and the colour transitions are
take into account within 2 strategies:
instantaneously, through perturbation theory, leading to a Langevin equation, analogous to QED \cite{Blaizot:2015hya}; or
as collisions, leading to a Botzman equation.
Moreover, in \cite{Gossiaux:2016htk}
an interesting framework, although not derived from first QCD principles, leads to a 
Schr\"{o}dinger-Langevin equation where the QCD features are taken into account within the parameters.

In order to develop a phenomenological approach, in Ref. \cite{Krouppa:2017jlg,Krouppa:2018lkt}
a lattice QCD vetted in-medium heavy-quark potential with anisotropic hydro QGP is introduced.
This in-medium potential achieves complex values at high temperatures.
Discrete values of the potential are obtained from lattice QCD.
A single T-dependent parameter remains, the Debye mass $m_D$.
The imaginary part, present in the lattice-vetted potential, leads to an easier dissociation of the $\Upsilon$ states when compare to 
the suppression obtained from a perturbative potential \cite{Krouppa:2016jcl}, thus introducing the possibility of recombination

Another dynamical in-medium transport model \cite{Yao:2017fuc}, based on 
perturbative NRQCD in a thermal QGP, shows that the
stochastic Boltzmann equations lead to heavy-quark
diffusion in the medium, which is necessary for the system to reach equilibrium. This model predicts the existence of flow $v_2$ 
from recombination.

\section{Conclusions}
In order to summarize, I would say that a modern approach 
to a theoretical description of quarkonium evolution in a QGP needs to take into account the following elements:
lattice QCD, effective field theory and open quantum system approach.
Taking only the 2 first would lead to a static framework --even if the expansion of the plasma is take into account--, described in terms of a real 
potential and spectral functions, which could reproduce the colour screening effect alone.
On the contrary, the introduction of a dynamical framework
through the OQS approach, permits the introduction of a static complex 
potential
which leads to a master equation 
from stochastic process, with time evolution of the heavy-quark states.
In this case, the effects of colour screening, Landau damping and singlet-octet transitions can be incorporated.

Note also that nuclear effects appear already in proton-nucleus collisions, where, in principle, no QGP formation is expected. The extrapolation of those effects to nucleus-nucleus collisions remains a matter of debate. 
In any case, the amount of some of them already in $pA$ collisions introduces questions about the nature of the medium in $pA$ but also in $AA$ collisions.

\noindent
{\small
{\bf Acknowledgments:} I thank Laboratoire LePrince-Ringuet, Ecole Polytechnique, Universite Paris-Saclay 
and the Ministerio de Ciencia e Innovacion of Spain under projects FPA2014-58293-
C2-1-P and Unidad de Excelencia Maria de Maetzu MDM-2016-0692 for
financial support. I also thank the organisers for inviting me to give this talk.}

\bibliographystyle{elsarticle-num}
\bibliography{QM2018nupha-Ferreiro}

\end{document}